\documentclass[aps,pra,twocolumn,groupedaddress,showpacs]{revtex4}

\usepackage{graphics}

\begin{document}

\title{Analytic description of atomic interaction at ultracold 
	temperatures: the case of a single channel}

\author{Bo Gao}
\email[Email: ]{bo.gao@utoledo.edu}
\homepage[Homepage: ]{http://bgaowww.physics.utoledo.edu}
\affiliation{Department of Physics and Astronomy,
	University of Toledo, MS 111,
	Toledo, Ohio 43606}

\date{\today}

\begin{abstract}

We present analytic descriptions of atomic interaction at 
ultracold temperatures
using both single-channel and multichannel quantum-defect theories.
In the case of a single channel, addressed in this paper, the expansion of 
[B. Gao, Phys. Rev. A \textbf{58}, 4222 (1998)]
is generalized to higher orders for angular momentum $l\ge 2$ to give a more complete
description of ultracold scattering, including an analytic description of
ultracold shape resonances of arbitrary $l$.
We also introduce a generalized scattering length that is well defined and useful for all
partial waves, to replace the traditional definition that fails for $l\ge 2$
due to the long-range van der Waals interaction. The results are used in
a companion paper to derive analytic descriptions of atomic interaction
around a magnetic Feshbach resonance of arbitrary angular momentum $l$.

\end{abstract}


\maketitle

\section{Introduction}

One of the key factors that has made cold-atom physics such a thriving
field has been the tunability of atomic interaction via Feshbach,
especially magnetic Feshbach resonances \cite{tie93,koh06,chi08}.
While the $s$ wave Feshbach resonances have received
most of the attention for many years, the same concept
is of course equally applicable
to any other, nonzero partial waves.
A number of such resonances have been observed experimentally 
\cite{reg03b,tic04,zha04,sch05,gae07,chi08}, including a recent observation
of a Feshbach resonance of $l=8$ \cite{kno08}. 
Before we can understand questions such as what happens to
a few-atom or a many-atom quantum system when a $p$ wave or a $d$ wave 
Feshbach resonance
is tuned around the threshold, we need first to understand 
the corresponding two-atom system, namely, how to describe atom-atom
scattering and atom-atom bound state with a Feshbach resonance of 
angular momentum $l$ around the threshold. 
This is the subject of
this and a companion study.

Recall that for the $s$ wave, a magnetic Feshbach resonance can be 
conveniently described by \cite{moe95}
\begin{equation}
a_{l=0}(B) = a_{bgl=0}\left(1-\frac{\Delta_{Bl=0}}{B-B_{0l=0}}\right) \;.
\label{eq:as1}
\end{equation}
Here $a_{l=0}$ represents the $s$ wave scattering length, which is tunable
by the magnetic field, $B$, around a Feshbach resonance, 
$a_{bgl=0}$ is a background scattering length, $\Delta_{Bl=0}$
is a measure of the width of the resonance, and $B_{0l=0}$ 
is the magnetic field at which 
$a_{l=0}=\infty$, corresponding to having a quasibound $s$ state right
at the threshold.
Once the scattering length is determined, the scattering properties
above the threshold, or the binding energy of an atom pair with large
and positive scattering lengths, can be determined, at least to a 
degree \cite{koh06,chi08}, 
from the effective-range theory (ERT) \cite{sch47,bla49,bet49}, which,
for positive energies, corresponds to the expansion 
\begin{equation}
k^{2l+1}\cot\delta_l=-\frac{1}{{a}_l}
	+\frac{1}{2}{r}_{el}k^2 +\cdots \;,
\label{eq:ere}	
\end{equation}
where $r_{el}$ is called the effective range.

The situation for $l\neq 0$ is considerably more complex due to the
long-range van der Waals interaction, $-C_6/r^6$, between atoms.
First, we have the obvious problem that the scattering length 
is not defined for $l\ge 2$ \cite{lev63,gao98b}, meaning that the energy
dependence of the scattering phase shift will necessarily differ from
that implied by Eq.~(\ref{eq:ere}). 
Second, a Feshbach resonance of $l\neq 0$,
when slightly above the threshold, is in fact a Feshbach/shape
resonance, the atoms in such a state see the angular-momentum
barrier just like they would in a single channel shape
resonance state.
When such a Feshbach resonance is tuned from above to 
below the threshold, its width, due to the presence of the
angular momentum barrier, can be expected to become increasingly
narrow and approaches zero when it crosses the threshold to 
become a true bound state.
Any complete theory for a Feshbach resonance of $l\neq 0$ has to
be able to describe this increasingly rapid energy variation as it
approaches the threshold. 

It is thus not surprising that the key to understanding
Feshbach resonances of $l\neq 0$, to be presented in a companion paper, 
turns out to be
the understanding of single channel shape resonances 
at ultracold energies, or equivalently, the threshold behaviors of
single channel shape resonances, which is addressed in detail in
this work. We point out that this is a nontrivial problem 
that, to the best of our knowledge, 
has not been done except for the $p$ wave description in
Ref.~\cite{gao98b}. 
Part of the difficulty can be attributed to the breakdown 
of the semiclassical approximation around 
the threshold \cite{wil45,kir78,jul89,gri93,tro98a,boi98,tro98b,fla99,gao99b,gao08a},
which implies, in particular, that the tunneling amplitude or 
probability through the angular momentum barrier \cite{gao08a}
cannot be obtained semiclassically.

Our analytic description of ultracold shape resonances 
is based on the small energy expansion of the quantum-defect theory (QDT) 
for $-1/r^6$ type of interactions \cite{gao98b,gao01,gao08a}.
It is presented here
in a general analytic framework for ultracold atomic interaction
that we call the QDT expansion, which is
applicable with or without the presence of ultracold shape resonances.
The presentation is organized as follows. 
In Sec.~\ref{sec:singlech}, we generalize the single-channel QDT expansion 
of Ref.~\cite{gao98b} to higher orders for $l\ge 2$ to give us a more complete
description of atom-atom scattering around the threshold, including an analytic
description of the threshold behaviors of single-channel shape resonances.
In Sec.~\ref{sec:schshape}, we present further understanding of ultracold
shape resonances by extracting from the QDT expansion analytic formulas
for their position, width, and background.
In Sec.~\ref{sec:gere}, we derive from the QDT expansion
a generalized effective-range expansion,
from which we introduce the concepts of a generalized scattering length and
a generalized effective range that are
well defined for all angular momenta $l$.
In Sec.~\ref{sec:schbsp}, we summarize the QDT expansions,
derived previously in Ref.~\cite{gao04c},
for the binding energies of the least-bound molecular state of
arbitrary angular momentum $l$, using the standardized notations of
Ref.~\cite{gao08a} that we adopt here. We also present 
in this section a few intermediate 
results that will be useful in studying magnetic Feshbach resonances
of arbitrary $l$. Further comments and remarks on the theory
are presented in Sec.~\ref{sec:discuss},
with conclusions given in Sec.~\ref{sec:conclude}.
The appendix presents analytic results of generalized scattering lengths
and other QDT parameters, for arbitrary $l$, 
for two types of model potentials, a hard sphere with an attract tail of the type
of $-C_6/r^6$, and the Lennard-Jones potential of the type of LJ(6,10).

The analytic description of atomic interaction around a 
magnetic Feshbach resonance of arbitrary $l$, which is
necessarily multichannel in nature \cite{tie93,koh06,chi08}, 
is developed in a companion paper. 
It will be accomplished through first, a rigorous
reduction of the underlying multichannel problem, as described by 
the multichannel quantum-defect theory (MQDT) for $-1/r^6$ type 
of interactions \cite{gao05a}, to an effective
single-channel problem, and a subsequent application of the
results of this work.

\section{QDT expansion for single-channel scattering}
\label{sec:singlech}

In this section, we derive and discuss the QDT expansion for
single-channel scattering. It differs considerably from the 
ERT \cite{sch47,bla49,bet49}, which assumes that
$k^{2l+1}\cot\delta_l$ is both an analytic \textit{and} a slowly
varying function of energy around the threshold, and can therefore
can be approximated by the first few terms in its energy expansion.
The QDT expansion makes no such assumptions.
The only quantities expanded are the universal functions 
of a scaled energy that are associated
with the long-range potential \cite{gao08a}.
There is no assumption about how $k^{2l+1}\cot\delta_l$ or
$\tan\delta_l$ may vary with energy.
It is partly for this reason that the QDT expansion can give analytic
description of an ultracold shape resonance, which has rapid
energy dependence that would have required, at least, partial
summation over all orders of energy in more standard approaches.
The QDT expansion is also more than a small-energy expansion.
It is simultaneously a large-$l$ expansion, in the sense that
for any fixed energy, there is a sufficiently large $l$ beyond which
it becomes applicable. As a related consequence, the energy range
over which the QDT expansion is applicable increases
rapidly for larger $l$.

While the derivation of the QDT expansion is somewhat tedious, 
the end result will be very simple. It is given by a single analytic formula
applicable to all angular momentum $l$ and regardless of whether
there is, or is not, an ultracold shape resonance.
To make the derivation easier to follow,
we first rewrite, in Sec.~\ref{sec:sch1}, the QDT for single 
channel scattering \cite{gao98b,gao01,gao08a}
in a form that makes its subsequent expansion, presented and
discussed in Sec.~\ref{sec:qdtexp},
fully transparent.

\subsection{QDT for single-channel scattering}
\label{sec:sch1}

For a single channel problem with long-range $-C_6/r^6$ interaction,
the scattering above the threshold is described rigorously in the
QDT by the following equation for the $K$ matrix \cite{gao98b,gao01,gao08a}
\begin{equation}
K_l = \tan\delta_l = ( Z^{c}_{gc}K^{c}-Z^{c}_{fc} )
	(Z^{c}_{fs} - Z^{c}_{gs}K^{c})^{-1} \;.
\label{eq:Kphy}	
\end{equation}
Here $K^c(\epsilon,l)$ is a short-range $K$ matrix that depends weakly 
on both the energy $\epsilon$
and the angular momentum $l$ \cite{gao08a}. 
The $Z^{c}_{xy}$ are elements of the $Z^c$ matrix for the $-1/r^6$ 
type of potentials, labeled here using the standardized notation of
Ref.~\cite{gao08a}. They are given explicitly by
\begin{eqnarray}
Z^{c}_{fs} &=& A^Z_l\left\{ \left[1-(-1)^l M_{\epsilon_s l}\tan\pi(\nu-\nu_0)
	\right]\sin(\pi\nu/2)X_l \right.\nonumber\\
	& & + \left.\left[1+(-1)^l M_{\epsilon_s l}\tan\pi(\nu-\nu_0)
	\right]\cos(\pi\nu/2)Y_l\right\} , \\
Z^{c}_{fc} &=& A^Z_l\left\{ \left[\tan\pi(\nu-\nu_0)-(-1)^l M_{\epsilon_s l}
	\right]\sin(\pi\nu/2)X_l \right.\nonumber\\
	& & + \left.\left[\tan\pi(\nu-\nu_0)+(-1)^l M_{\epsilon_s l}
	\right]\cos(\pi\nu/2)Y_l\right\} ,\\
Z^{c}_{gs} &=& A^Z_l\left\{ \left[1+(-1)^l M_{\epsilon_s l}\tan\pi(\nu-\nu_0)
	\right]\cos(\pi\nu/2)X_l \right.\nonumber\\
	& & - \left.\left[1-(-1)^l M_{\epsilon_s l}\tan\pi(\nu-\nu_0)
	\right]\sin(\pi\nu/2)Y_l\right\} , \\
Z^{c}_{gc} &=& A^Z_l\left\{ \left[\tan\pi(\nu-\nu_0)+(-1)^l M_{\epsilon_s l}
	\right]\cos(\pi\nu/2)X_l \right.\nonumber\\
	& & - \left.\left[\tan\pi(\nu-\nu_0)-(-1)^l M_{\epsilon_s l}
	\right]\sin(\pi\nu/2)Y_l\right\} ,
\end{eqnarray}
where
\[
A^Z_l = \frac{G_{\epsilon_s l}(\nu)\cos\pi(\nu-\nu_0)}
	{\sqrt{2}(X_l^2+Y_l^2)\sin\pi\nu} \;,
\]
$\nu_0 = (2l+1)/4$, $M_{\epsilon_s l}=G_{\epsilon_s l}(-\nu)/G_{\epsilon_s l}(\nu)$,
with the characteristic exponent $\nu$, and functions $X_l$, $Y_l$, and 
$G_{\epsilon_s l}$ being defined in Ref.~\cite{gao98a}. 
They are all universal functions of a scaled energy,
\begin{equation}
\epsilon_s = \epsilon/s_E \;,
\label{eq:es}
\end{equation}
where $s_E = (\hbar^2/2\mu)(1/\beta_6)^2$ is the energy scale  
corresponding to the length scale $\beta_6=(2\mu C_6/\hbar^2)^{1/4}$ that is
associated with the $-C_6/r^6$ potential. For the purpose of 
providing orders of magnitudes, these scales, and also the related 
time scale, $s_T=\hbar/s_E$,
are tabulated in Table~\ref{tb:scales} for selected alkali-metal dimers.
They are all determined by the $C_6$ coefficient and the reduced mass $\mu$.
\begin{table}
\caption{Sample scale parameters for $A+A$ type of systems where $A$ is
an alkali-metal atom.
The $\beta_6=(2\mu C_6/\hbar^2)^{1/4}$ is the length scale. 
The $s_E=(\hbar^2/2\mu)(1/\beta_6)^2$ 
is the corresponding energy scale. It is given both in units of $\mu$K and in
units of MHz. $s_T=\hbar/s_E$ is the corresponding time scale. 
All are determined by the $C_6$ coefficient and atomic masses. 
\label{tb:scales}}
\begin{ruledtabular}
\begin{tabular}{cccccc}
Atom & $C_6$ (a.u.) & $\beta_6$ (a.u.) & $s_E/k_B$ ($\mu$K) & $s_E/h$ (MHz) & $s_T$ (ns) \\
\hline
$^6$Li  & 1393.39\footnotemark[1] & $62.52$ 
	& $7368$ & $153.5$ & $1.037$ \\
$^{23}$Na  & $1556$\footnotemark[2] & $89.86$ 
	& $933.1$ & $19.44$ & $8.186$ \\
$^{40}$K  & 3897\footnotemark[2] & $129.8$ 
	& $257.3$ & $5.360$ & $29.69$ \\
$^{85}$Rb  & $4707$\footnotemark[3] & $164.3$ 
	& $75.58$ & $1.575$ & $101.1$ \\
$^{133}$Cs  & 6860\footnotemark[4] & $201.9$ 
	& $31.97$ & $0.6662$ & $238.9$
\end{tabular}
\end{ruledtabular}
\footnotetext[1]{From Ref.~\cite{yan96}}
\footnotetext[2]{From Ref.~\cite{der99}}
\footnotetext[3]{From Refs.~\cite{mar02,cla03}}
\footnotetext[4]{From Ref.~\cite{chi04}}
\end{table}
Equation~(\ref{eq:Kphy}) gives an exact description of single-channel
atomic scattering in terms of a set of universal functions determined solely
by the long-range interaction.
All the short-range physics are encapsulated in $K^c(\epsilon, l)$,
which is slowly varying in both $\epsilon$ and $l$ around the threshold
\cite{gao08a}.

Instead of the $K^c$ parameter, the short-range atomic interaction 
in any partial wave can also be described using alternative parameters 
such as the quantum-defect $\mu^c$ or the parameter $K^{c0}_l$ \cite{gao08a}.
They are all related, but have different utilities for different
purposes. For this work, which focuses on the ultracold regime around the
threshold, the most convenient parameter is $K^{c0}_l$.
It is defined as the short-range $K$ matrix 
associated with the $f^{c0}$ and $g^{c0}$ reference pair 
of Ref.~\cite{gao08a}, and is related, for $-1/r^6$ type of 
potentials, to the $K^c$ and $\mu^c$ by
\begin{eqnarray}
K^{c0}_l(\epsilon) &=& \frac{K^c(\epsilon, l)-\tan(\pi\nu_0/2)}
	{1+\tan(\pi\nu_0/2)K^c(\epsilon, l)} \;,
\label{eq:Kc0l} \\
	&=&\tan[\pi\mu^c(\epsilon, l)-l\pi/4]\;.
\end{eqnarray}
With this definition, $K^{c0}_l(\epsilon=0)=0$ corresponds to having a bound or
quasibound state of angular momentum $l$ right at the threshold, 
a small and positive $K^{c0}_l(0)$ corresponds to having a bound
state of $l$ close to the threshold,
and a small and negative $K^{c0}_l(0)$ corresponds to having a 
shape resonance of angular momentum $l$ close to the threshold \cite{gao00,gao04b,gao04c}.
This property of $K^{c0}_l$, which was called $x_l$ in Refs.~\cite{gao04b,gao04c}
and is related to the $K^0_l$ of Ref.~\cite{gao98b} by $K^{c0}_l=-K^0_l$,
allows it to be used as an expansion parameter when there is
a state close to the threshold \cite{gao04c}. 
As will become clear later in the paper, $K^{c0}_l$ is also the
short-range parameter that has the simplest relation to the 
scattering lengths and the generalized scattering lengths.

Defining $\theta_l$ by
\begin{eqnarray}
\sin\theta_l &=& Y_l(X_l^2+Y_l^2)^{-1/2} \;,\\
\cos\theta_l &=& X_l(X_l^2+Y_l^2)^{-1/2} \;,
\end{eqnarray}
the elements of the $Z^{c}$ matrix can be further written as
\begin{eqnarray}
Z^{c}_{fs} &=& B^Z_l 
	\left[\sin\left(\frac{1}{2}\pi\nu+\theta_l\right) \right.\nonumber\\
	& &\left.-(-1)^l M_{\epsilon_s l}\tan\pi(\nu-\nu_0)
		\sin\left(\frac{1}{2}\pi\nu-\theta_l\right)\right] \;, 
\label{eq:Zcfs}	\\
Z^{c}_{fc} &=& B^Z_l \left[\tan\pi(\nu-\nu_0)
	\sin\left(\frac{1}{2}\pi\nu+\theta_l\right) \right.\nonumber\\
	& &\left.-(-1)^l M_{\epsilon_s l}
	\sin\left(\frac{1}{2}\pi\nu-\theta_l\right)\right] \;, \\
Z^{c}_{gs} &=& B^Z_l 
	\left[\cos\left(\frac{1}{2}\pi\nu+\theta_l\right) \right.\nonumber\\
	& &\left.+(-1)^l M_{\epsilon_s l}\tan\pi(\nu-\nu_0)
		\cos\left(\frac{1}{2}\pi\nu-\theta_l\right)\right] \;, \\
Z^{c}_{gc} &=& B^Z_l \left[\tan\pi(\nu-\nu_0)
	\cos\left(\frac{1}{2}\pi\nu+\theta_l\right) \right.\nonumber\\
	& &\left.+(-1)^l M_{\epsilon_s l}
	\cos\left(\frac{1}{2}\pi\nu-\theta_l\right)\right] \;,
\label{eq:Zcgc}	
\end{eqnarray}
where
\begin{equation}
B^Z_l = \frac{G_{\epsilon_s l}(\nu)\cos\pi(\nu-\nu_0)}
	{\sqrt{2}(X_l^2+Y_l^2)^{1/2}\sin\pi\nu} \;.
\end{equation}
Substituting Eqs.~(\ref{eq:Zcfs})-(\ref{eq:Zcgc}) into Eq.~(\ref{eq:Kphy}) and
using $K^{c0}_l(\epsilon)$ as the short-range parameter, we can rewrite the
QDT equation for the $K$ matrix as 
\begin{widetext}
\begin{equation}
K_l = \tan\delta_l = -\tan[\pi(\nu-\nu_0)]
	-\widetilde{A}_{ls}(\epsilon_s)k_s^{2l+1}
	\frac{1-\tan^2[\pi(\nu-\nu_0)]}
	{1+\widetilde{A}_{ls}(\epsilon_s)k_s^{2l+1}\tan[\pi(\nu-\nu_0)]} \;,
\label{eq:Kl}	
\end{equation}
where $k_s = \epsilon_s^{1/2}=k\beta_6$, and
\begin{equation}
\widetilde{A}_{ls}(\epsilon_s) = 
	\left[\frac{(-1)^lM_l\sin\pi\nu_0}{k_s^{2l+1}}\right]
	\frac{[1+(-1)^lK^{c0}_l]\{1+\tan\theta_l\tan[\pi(\nu-\nu_0)/2]
	-[(-1)^l-K^{c0}_l]\{\tan\theta_l-\tan[\pi(\nu-\nu_0)/2]\}}
	{K^{c0}_l-\tan\theta_l-\tan[\pi(\nu-\nu_0)/2]-K^{c0}_l\tan\theta_l\tan[\pi(\nu-\nu_0)/2]} \;.
\label{eq:tAl}	
\end{equation}
\end{widetext}
This is still an exact expression for $K_l$, and it has been written in a way
to make its QDT expansion fully transparent.

\subsection{QDT expansion}
\label{sec:qdtexp}

For small energies around the threshold, or for arbitrary energy but
sufficiently large $l$,
the quantities in Eq.~(\ref{eq:Kl}) can be represented
by expansions \cite{gao04c} that derive
straightforwardly from the analytic solution for
the $-1/r^6$ type of potential \cite{gao98a}:
\begin{equation}
\nu-\nu_0 = -\frac{3}{2^5\nu_2(\nu_2^2-1)(\nu_2^2-4)}\epsilon_s^2 
	+O(\epsilon_s^4)\;,
\label{eq:nuexp}	
\end{equation}
\begin{equation}
\tan\theta_l = -\frac{1}{2^2(\nu_2^2-1)}\epsilon_s +O(\epsilon_s^3)\;,
\label{eq:thetaexp}
\end{equation}
\begin{widetext}
\begin{eqnarray}
M_{\epsilon_s l} &=& (-1)^l\frac{\pi^2}{2^{4\nu_2-1}\sin(\pi\nu_0)}
	\frac{1}{[\Gamma(\nu_0)\Gamma(\nu_2+1)]^2}|\epsilon_s|^{\nu_2} \nonumber\\
	& &\times\left\{1+2(\nu-\nu_0)\ln|\epsilon_s|
		-\left[(-1)^l\pi+8\ln 2+\frac{4}{\nu_2}
		+2\psi(\nu_0)\right](\nu-\nu_0)\right.\nonumber\\
	& &\left.+\frac{1}{2^6\nu_2}\left[\frac{1}{(\nu_2+1)^2(\nu_2+2)}
		-\frac{16}{(\nu_2^2-4)^2}-\frac{1}{(\nu_2-1)^2(\nu_2-2)}
		\right]\epsilon_s^2\right\} \nonumber\\ 
	& &+O\left[|\epsilon_s|^{\nu_2+4}(\ln|\epsilon_s|)^2\right] \;,
\label{eq:Mexp}	
\end{eqnarray}
\end{widetext}
where $\nu_0 = (2l+1)/4$, as defined earlier, $\nu_2\equiv 2\nu_0 = l+1/2$, 
and $\psi(x)$ is the digamma function \cite{abr64}. 

Combining these expansions with Eqs.~(\ref{eq:Kl}) and (\ref{eq:tAl}) shows
that the $K$ matrix has the following structure
\begin{widetext}
\begin{equation}
\tan\delta_l= -\pi(\nu-\nu_0)
	-\bar{a}_{sl}k_s^{2l+1}
	\frac{1+(-1)^lK^{c0}_l-[(-1)^l-K^{c0}_l][\theta_l-\pi(\nu-\nu_0)/2]+O(k_s^6)}
	{K^{c0}_l-\theta_l-\pi(\nu-\nu_0)/2+O(k_s^6)}
	\left[1+O(k_s^4\ln k_s)\right] + O(k_s^8)\;,
\label{eq:Klexp}	
\end{equation}
\end{widetext}
where
\begin{eqnarray}
\bar{a}_{sl} &=& \frac{\pi^2}{2^{4l+1}[\Gamma(l/2+1/4)\Gamma(l+3/2)]^2} \nonumber\\
	&=& \frac{\pi}{2^{2l-1}[\Gamma(l/2+1/4)]^2[(2l+1)!!]^2} \;,
\end{eqnarray}
is what we call the scaled mean scattering length for angular momentum $l$. 
It is a generalization of the $s$
wave mean scattering length of Gribakin and Flambaum \cite{gri93} 
to $p$ \cite{gao04c} and higher partial waves. More explicitly,
\begin{equation}
\bar{a}_{sl=0} = \frac{2\pi}{[\Gamma(1/4)]^{2}} \approx 0.4779888 \;,
\end{equation}
\begin{equation}
\bar{a}_{sl=1} = \frac{[\Gamma(1/4)]^{2}}{36\pi} \approx 0.1162277 \;.
\end{equation}
For larger $l$, it decreases rapidly and can be either computed 
directly or obtained from $\bar{a}_{sl=0}$ and $\bar{a}_{sl=1}$ 
using the recurrence relation
\begin{equation}
\bar{a}_{sl+2} = \frac{1}{[(2l+5)(2l+3)(2l+1)]^2}\bar{a}_{sl} \;,
\end{equation}
which gives, e.g., $\bar{a}_{sl=2} = \bar{a}_{sl=0}/225$.
Together, they represent a set of universal numbers determined solely 
by the long-range potential. The corresponding mean
scattering length, with scale included, 
is defined as $\bar{a}_l=\bar{a}_{sl}\beta_6^{2l+1}$.

Equation~(\ref{eq:Klexp}), if kept to the order of $k_s^5$, gives 
the following result for the $s$ wave
\begin{widetext}
\begin{equation}
\tan\delta_{l=0} \approx \frac{\pi}{15}k_s^4
	-\bar{a}_{sl=0}k_s
	\frac{1+K^{c0}_l-(1-K^{c0}_l)\left(\frac{1}{3}k_s^2+\frac{\pi}{30}k_s^4\right)}
	{K^{c0}_l-\frac{1}{3}k_s^2+\frac{\pi}{30}k_s^4}
	\left[1-\frac{4}{15}k_s^4\ln k_s
	+\frac{2}{15}\left(\frac{22}{5}+\ln 2 
	-\gamma\right)k_s^4\right] \;,
\label{eq:Ks}
\end{equation}
\end{widetext}
where $\gamma=0.5772156649\dots$ is the Euler's constant \cite{abr64}. 
Results of even higher orders
are possible. An $s$ wave result to the order of $k_s^6$ was given 
in Ref.~\cite{gao98b}. Similar results can be written down for other partial waves. 

For the range of energies of interest in cold-atom physics, this level of
complexity is actually unnecessary, especially for higher partial
waves. For simplicity and easier application, 
we will use the following approximation of Eq.~(\ref{eq:Klexp}) for all $l$
\begin{equation}
\tan\delta_l \approx K^{(B)}_l+K^{(D)}_l \;,
\label{eq:qdtexp1}	
\end{equation}
where
\begin{eqnarray}
K^{(B)}_l &\approx& -\pi(\nu-\nu_0) \nonumber\\
	&\approx& \frac{3\pi}{(2l+5)(2l+3)(2l+1)(2l-1)(2l-3)}\epsilon_s^{2} \;,
\label{eq:qdtexp2}	
\end{eqnarray}
is a term that could have been derived from the 
Born approximation (see, e.g., Ref.~\cite{lan77}),
and
\begin{equation}
K^{(D)}_l \approx -\widetilde{A}_{sl}(\epsilon_s)k_s^{2l+1}\;,
\label{eq:qdtexp3}	
\end{equation}
describes the deviation from the Born term. Here
\begin{eqnarray}
\widetilde{A}_{sl}(\epsilon_s) &\approx& \bar{a}_{sl}\left[
	(-1)^l+\frac{1+K^{c0}_l\theta_l}
	{K^{c0}_l-\theta_l-\pi(\nu-\nu_0)/2}\right] \;,
\label{eq:qdtexp4} \\
	&=& \bar{a}_{sl}\left[
	(-1)^l+\frac{(2l+3)(2l-1)-K^{c0}_l\epsilon_s}
	{(2l+3)(2l-1)K^{c0}_l+\epsilon_s+w_l\epsilon_s^2}\right] \;,
\label{eq:qdtexp5}	
\end{eqnarray}
can be regarded as a scaled, energy-dependent, generalized scattering length. 
Both expressions for $\widetilde{A}_{sl}(\epsilon_s)$ are useful for different
purposes. In Eq.~(\ref{eq:qdtexp5}), $w_l$ is an $l$-dependent constant 
defined by
\begin{equation}
w_l=\frac{3\pi}{2(2l+5)(2l+1)(2l-3)} \;.
\label{eq:wl}
\end{equation}
Equation~(\ref{eq:qdtexp4}), while less explicit compared to
Eq.~(\ref{eq:qdtexp5}), is more convenient for a number
of conceptual purposes.
Here $\pi(\nu-\nu_0)$ is given by Eq.~(\ref{eq:qdtexp2}), and
$\theta_l$ is given by Eq.~(\ref{eq:thetaexp}), or more explicitly by
\begin{equation}
\theta_l \approx -\frac{1}{(2l+3)(2l-1)}\epsilon_s \;.
\label{eq:thetal}
\end{equation}

\begin{table}
\caption{Critical scaled energies, $\epsilon_{scl}$, for different angular
	momentum $l$. Beyond $\epsilon_{scl}$, the characteristic
	exponent $\nu$ for the $-1/r^6$ solutions \cite{gao98a}
	moves off the real axis and becomes complex.}
\label{tb:esc}
\begin{ruledtabular}
\begin{tabular}{cccc}
$l$ & $\epsilon_{scl}$ & $l$ & $\epsilon_{scl}$ \\
\hline
0 & $1.544707$ & 6  & $94.82401$ \\
1 & $2.358067$ & 7  & $212.3067$ \\
2 & $6.891073$ & 8  & $206.8228$ \\
3 & $25.29322$ & 9  & $406.8301$ \\
4 & $33.17273$ & 10 & $383.5433$\\
5 & $89.85261$ & 11 & $687.4041$
\end{tabular}
\end{ruledtabular}
\end{table}
Together, Eqs.~(\ref{eq:qdtexp1})-(\ref{eq:qdtexp5}) give a
single analytic formula for low-energy atomic scattering that we
call the QDT expansion. It is applicable to all $l$, with or without
the presence of ultracold shape resonances. It has 
the following additional characteristics.
(a) The QDT expansion remains applicable no matter how rapidly 
the $K^{c0}_l(\epsilon)$ parameter may depend on energy.
In arriving at Eqs.~(\ref{eq:qdtexp1})-(\ref{eq:qdtexp5}), the only 
quantities that are expanded
are the universal functions associated with the long-range
potential. There is no assumption about the values of $K^{c0}_l(\epsilon)$, 
or how it may depend on energy.
While this feature of the QDT expansion is not important for true single channel
cases, for which the energy dependence of $K^{c0}_l$ is almost always
negligible \cite{gao98b,gao01}, 
it will become crucial when we apply it to effective single channel
problems that are derived from intrinsically multichannel ones,
for which the energy dependence of $K^{c0}_l$ is generally important.
(b) The energy range over which the QDT expansion is applicable 
increase rapidly with $l$, roughly as $l^3$. 
This is consistent with the fact that the QDT expansion is
simultaneously a large $l$ expansion \cite{gao04c}. For any energy, there is
a sufficiently large $l$ beyond which it becomes applicable.
The only condition for the applicability of the QDT expansion
is $\epsilon_s\ll \epsilon_{scl}$, where
$\epsilon_{scl}$ is the scaled critical energy beyond which the characteristic
exponent $\nu$ becomes complex \cite{gao98a}.
Values of $\epsilon_{scl}$ for the first twelve partial waves
are listed in Table~\ref{tb:esc}.
For large $l$, they correspond roughly to the scaled height of the 
angular momentum barrier, given for $-1/r^6$ type of potentials by 
$H_{sl}=(2/3^{3/2})[l(l+1)]^{3/2}$.
As will be discussed in more detail elsewhere, the critical energy 
is also the energy around which the behavior of the system goes 
from being quantum to being semiclassical \cite{fla99,gao08a}. 
Thus the QDT expansion can also be regarded as a
quantum expansion, applicable over a quantum 
region of energies where
the quantum reflection probability is close to 1 \cite{gao98a}.

\begin{figure}
\scalebox{0.4}{\includegraphics{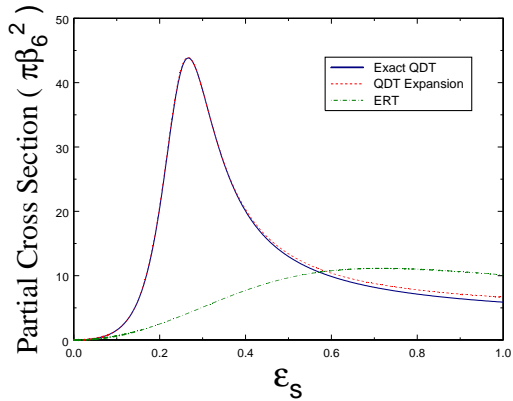}}
\caption{(Color online) $p$ wave partial cross sections for $a_{l=1}/\bar{a}_{l=1}=-20$
(corresponding to $K^{c0}_{l=1}\approx -0.0526$).
Results of the QDT expansion (dashed line) are compared to the exact 
QDT results computed from Eq.~(\ref{eq:Kphy}) (solid line),
and the results of ERT (dash-dot line).
\label{fig:pshape_m20}}
\end{figure}
\begin{figure}
\scalebox{0.4}{\includegraphics{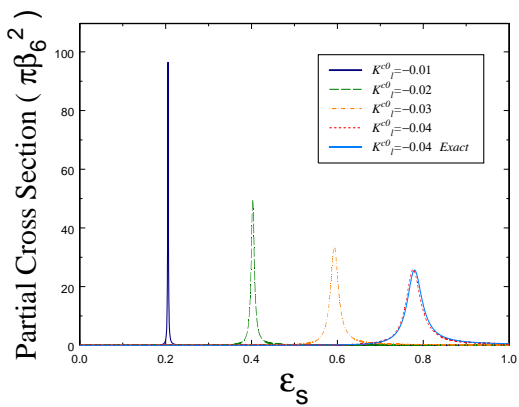}}
\caption{(Color online) $d$ wave shape resonances as described by the QDT expansion 
for different values of $K^{c0}_{l=2}$. 
The exact QDT results computed from Eq.~(\ref{eq:Kphy}) are only 
shown for $K^{c0}_{l=2}=-0.04$ as they
become indistinguishable from those of QDT expansion for smaller $|K^{c0}_{l=2}|$.
For higher partial waves, the QDT expansions are more accurate, 
and applicable over a greater range of scaled energies than shown here
for the $d$ wave.
\label{fig:dshape}}
\end{figure}
Figures~\ref{fig:pshape_m20} and \ref{fig:dshape} illustrate the QDT expansion
for $p$ and $d$ waves, respectively, using cases that have
a shape resonance in the threshold region. Since our interest here is
in the case of a single channel, $K^{c0}_l$ is taken to be a constant \cite{gao98b,gao01},
$K^{c0}_l=K^{c0}_l(\epsilon=0)$, which is related in a simple way
to the scattering length by Eq.~(\ref{eq:ga}) or (\ref{eq:Kc0ga}), 
to be discussed in more details later.
Note that the QDT expansion is applicable regardless of how narrow the
shape resonances may be. In fact it is more accurate for narrower resonances
which are necessarily located at smaller energies, as illustrated
in Fig.~\ref{fig:dshape}. For higher partial waves, the results of 
the QDT expansion are no longer distinguishable from the exact QDT
results, computed using Eq.~(\ref{eq:Kphy}), in the range of energies shown 
in Figs.~\ref{fig:pshape_m20} and \ref{fig:dshape}. 
Such results are easily calculated from the analytic formula, 
and are therefore not shown.

Figure~\ref{fig:pshape_m20} is also used to illustrate the limitation 
of the ERT \cite{sch47,bla49,bet49}. 
In the figure, the ERT results are computed from [see Eq.~(\ref{eq:ere})]
\begin{equation}
k^3\cot\delta_{l=1} = -1/a_{l=1} \;.
\label{eq:ertp}
\end{equation}
Note that since the effective range is not defined for the $p$ wave
due to the van der Waals interaction \cite{lev63,gao98b},
it is incorrect to add an effective range term to Eq.~(\ref{eq:ertp}).
The lowest order correction to the right-hand-side of Eq.~(\ref{eq:ertp})
is of the order $k$ (see Eqs.~(\ref{eq:qdtexp1})--(\ref{eq:qdtexp5}), or Ref.~\cite{gao98b}), 
not of the order of $k^2$ as implied by ERT.
Figure~\ref{fig:pshape_m20} clear illustrates the failure of ERT, 
which misses the $p$ wave shape resonance completely. 
The consequence of such a failure for two atoms
in a trap has been illustrated elsewhere \cite{che07}.
Careful readers should note that ERT description of $p$ wave interaction
is still sometimes incorrectly used in the literature. This is unfortunate, 
considering that a correct description, at least in the case of a single channel, 
has been available for some time \cite{gao98b}.

\begin{figure}
\scalebox{0.4}{\includegraphics{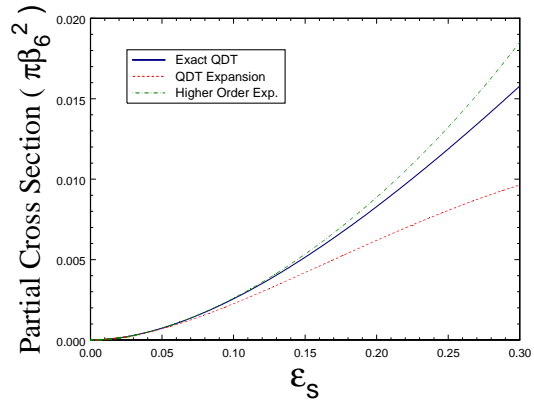}}
\caption{(Color online) $s$ wave partial cross sections for $a_{l=0}=0$, 
representing the worst-case scenario for the QDT expansion. 
The results of the recommended QDT expansion (dashed line) are compared 
to the exact QDT results from Eq.~(\ref{eq:Kphy}) (solid line), and
the results of a higher order expansion, given by Eq.~(\ref{eq:Ks})
(dash-dot line). 
\label{fig:swave_0}}
\end{figure}
Among all partial waves, the QDT expansion for the $s$ wave
has the smallest range of applicability because the critical scaled
energy, given in Table~\ref{tb:esc}, is smallest for $l=0$. 
For the $s$ wave, the most stringent test of the expansion 
is for $a_{l=0}=0$. It corresponds to a case where the lowest
order contribution, $-a_{l=0}k$, to $\tan\delta_l$ goes to zero, and we are directly
testing the higher order terms. As will be discussed in more detail in Sec.~\ref{sec:gere},
this is also the case where the effective range expansion, even in its
generalized form of Sec.~\ref{sec:gere}, fails completely.
Figure~\ref{fig:swave_0} illustrates the accuracy of the QDT expansion for
this worst case scenario. It shows that the simpler QDT expansion that we are
recommending here is accurate for roughly $\epsilon_s<0.1$.
From Table~\ref{tb:scales}, it is clear that this range of
scaled energies already
covers all energies of interest in cold-atom physics.
The QDT expansion is more accurate and applicable over a greater
range of energies in all other cases and for all other partial waves.

For the $s$ and $p$ partial waves, our results here are consistent with
those of Ref.~\cite{gao98b}, except they are now written in simpler
forms that are also better for physical interpretation.
For $l\ge 2$, the $K^{(D)}$ term, which is of the
order of $k_s^{2l+1}$ under nonresonant conditions of $|K^{c0}_l|$ being
of the order of 1 or greater, is normally negligible,
as was done in Ref.~\cite{gao98b}. In doing so, however, we missed
in our previous work \cite{gao98b} the 
analytic description of low-energy shape resonances for
$l\ge 2$, and an opportunity to define the generalized scattering
length and the generalized effective range. 
These subjects are addressed in the next two sections,
respectively.

\section{Threshold behavior of shape resonances}
\label{sec:schshape}

The QDT expansion of Eqs.~(\ref{eq:qdtexp1})-(\ref{eq:qdtexp5})
is applicable whether or not there is a shape resonance in the
threshold region.
When such a resonance does exist, which occurs for $l\ge 1$ and
a small and negative $K^{c0}_l$, namely for $K^{c0}_l<0$ and $|K^{c0}_l|\ll 1$,
further conceptual understanding can be achieved by extracting
from the QDT expansion the standard parameters characterizing
a resonance, namely its position, width, and background 
(see, e.g., Ref.~\cite{tay06}).

The position of a shape resonance in the threshold region 
can be determined from the root of
the denominator in Eq.~(\ref{eq:qdtexp3}), 
\begin{equation}
\theta_l+\pi(\nu-\nu_0)/2 = K^{c0}_l \;.
\label{eq:shapepos}
\end{equation}
It has solution only for a small and negative $K^{c0}_l$,
for which it can be solve perturbatively to give
the scaled resonance position as
\begin{equation}
\epsilon_{sl} \approx -(2l+3)(2l-1)K^{c0}_l
	\left[1+w_l(2l+3)(2l-1)K^{c0}_l\right] \;,
\label{eq:esl}
\end{equation}
in which $w_l$ is defined earlier by Eq.~(\ref{eq:wl}).
Around the shape resonance, 
the term $K^{(D)}_l=-\widetilde{A}_{sl}(\epsilon_s)k_s^{2l+1}$
can be written as
\begin{equation}
K^{(D)}_{l} \approx K^{(D)}_{bgl}(\epsilon_s)-\frac{1}{2}\frac{\gamma_{sl}}
	{\epsilon_s-\epsilon_{sl}} \;,
\label{eq:KDshape}	
\end{equation}
with the scaled width, $\gamma_{sl}$, given by
\begin{eqnarray}
\gamma_{sl} &\approx& 2[(2l+3)(2l-1)]^{l+3/2}\bar{a}_{sl}(-K^{c0}_l)^{l+1/2} \nonumber\\
	& &\times\left[1+\frac{3\pi(2l+3)(2l-1)}
	{4(2l+1)(2l-3)}K^{c0}_l\right] \;,
\label{eq:wsl1}	\\
	&\approx& \frac{2(2l+3)(2l-1)\bar{a}_{sl}(\epsilon_{sl})^{l+1/2}}
	{1+2w_l\epsilon_{sl}} \;,
\label{eq:wsl2}
\end{eqnarray}
and the contribution of $K^{(D)}_l$ to the background, $K^{(D)}_{bgl}(\epsilon_s)$,
given by
\begin{equation}
K^{(D)}_{bgl=1} \approx -\frac{5\bar{a}_{sl}}{1-\frac{\pi}{7}\epsilon_{sl}}
	\left(\frac{\epsilon_s^{3/2}-\epsilon_{sl}^{3/2}}
	{\epsilon_s-\epsilon_{sl}}\right)
	-\left(\frac{5\pi}{14}-1\right)\bar{a}_{sl}\epsilon_s^{3/2} \;,
\end{equation}
\begin{equation}
K^{(D)}_{bgl=2} \approx -\frac{21\bar{a}_{sl}}{1+\frac{\pi}{15}\epsilon_{sl}}
	\left(\frac{\epsilon_s^{5/2}-\epsilon_{sl}^{5/2}}
	{\epsilon_s-\epsilon_{sl}}\right) \;,
\end{equation}
\begin{equation}
K^{(D)}_{bgl>2} \approx 0 \;.
\label{eq:KDbgdg}
\end{equation}
Corresponding to the width of Eqs.~(\ref{eq:wsl1}) and (\ref{eq:wsl2}),
there is a well-defined lifetime (see, e.g., Ref.~\cite{tay06})
\begin{equation}
\tau_l = \tau_{sl}s_T = (1/\gamma_{sl})s_T\;,
\label{eq:taul}
\end{equation}
where $s_T=\hbar/s_E$ is the time scale associated with the length
scale $\beta_6$, with sample values for alkali-metal atoms given
in Table~\ref{tb:scales}, and $\tau_{sl} = \tau_l/s_T= 1/\gamma_{sl}$
is the scaled lifetime.

\begin{figure}
\scalebox{0.4}{\includegraphics{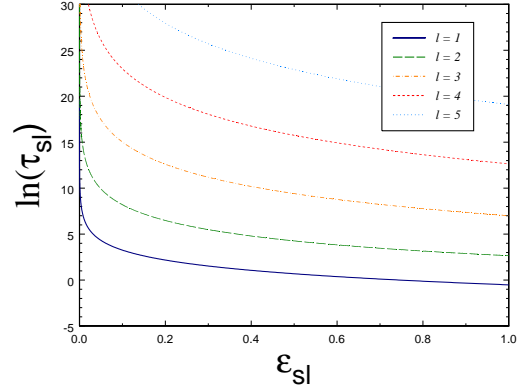}}
\caption{(Color online) Universal relation between the scaled lifetime and 
the scaled position of a single-channel shape resonance, as given by 
Eqs.~(\ref{eq:wsl2}) and (\ref{eq:taul}), for $-1/r^6$ type of long-range 
potentials. The lifetime can change by many orders of magnitude for a 
small change in the resonance position.
\label{fig:gamsl}}
\end{figure}
Equations~(\ref{eq:wsl1}) and (\ref{eq:wsl2}) show that the width of
a shape resonance goes to zero as it approaches the threshold, below which
it becomes a true bound state. The corresponding lifetime goes to infinity.
Equation~(\ref{eq:wsl2}) further shows that the scaled width, and therefore
the scaled lifetime, when viewed as a function of the scaled resonance
position,
follows a universal behavior that is uniquely determined by the long-range
interaction. In other words, while the resonance position itself,
as given by Eq.~(\ref{eq:esl}), depends on the short-range parameter $K^{c0}_l$,
the functional form of the scaled width or lifetime versus the scaled resonance position
is independent of it. This universal behavior is illustrated in
Figures~\ref{fig:gamsl} for the first few partial waves.
It also illustrates that the lifetime of a shape resonance
can change by many orders of magnitudes,
from a microscopic scale of $s_T$ (see Table~\ref{tb:scales}), to a macroscopic
scale of seconds or longer, as the shape resonance approaches the threshold.
The existence of such a potentially macroscopic time scale \cite{kno08}
is one of the key differences between the coupling of
atoms in $l\neq 0$ and $l=0$ partial waves.

\section{Generalized scattering length and generalized effective range}
\label{sec:gere}

The QDT expansion, Eqs.~(\ref{eq:qdtexp1})-(\ref{eq:qdtexp5}), 
already suggests that one may be
able to define a generalized scattering length for an arbitrary $l$. This, and
the definition of a generalized effective range for an arbitrary $l$, 
can be done in a way that more closely resembles to the standard 
ERT \cite{sch47,bla49,bet49}, as follows.

Defining $\delta^{(B)}_l\equiv -\pi(\nu-\nu_0)$, we have from 
Eqs.~(\ref{eq:Kl})-(\ref{eq:Mexp})
\begin{widetext}
\begin{equation}
k_s^{2l+1}\cot(\delta_l-\delta^{(B)}_l)=
	-\left(\frac{1}{\bar{a}_{sl}}\right)
	\frac{K^{c0}_l-\theta_l+O(k_s^4)}
	{1+(-1)^lK^{c0}_l-[(-1)^l-K^{c0}_l]\theta_l+O(k_s^4)}
	 \left[1+O(k_s^4\ln k_s)\right] + O(k_s^4)\;.
\label{eq:cotexp}	 
\end{equation}
\end{widetext}
In the case of a single channel, the energy dependence of $K^{c0}_l$
is completely negligible \cite{gao98b,gao01}. A further expansion of the denominator
in Eq.~(\ref{eq:cotexp}) gives the following generalized effective 
range expansion
\begin{equation}
k^{2l+1}\cot(\delta_l-\delta^{(B)}_l)=-\frac{1}{\widetilde{a}_l}
	+\frac{1}{2}\widetilde{r}_{el}k^2 +O(k^4\ln k) \;,
\label{eq:gere}	
\end{equation}
with the generalized scattering length, $\widetilde{a}_l$, given by
\begin{equation}
\widetilde{a}_l = \bar{a}_{l}\left[(-1)^l+\frac{1}{K^{c0}_l(\epsilon=0)}\right] \;,
\label{eq:ga}
\end{equation}
where $\bar{a}_{l}=\bar{a}_{sl}\beta_6^{2l+1}$ is the mean scattering
length for angular momentum $l$ that we have defined earlier 
(with scale included). The
generalized effective range, $\widetilde{r}_{el}$, is given by
\begin{equation}
\widetilde{r}_{el} = -\frac{2\bar{a}_{l}\beta_6^2}{(2l+3)(2l-1)\widetilde{a}_{l}^2}
	\left\{1+\left[(-1)^l-(\widetilde{a}_{l}/\bar{a}_{l})\right]^2\right\} \;.
\label{eq:gr}	
\end{equation}
For quantum systems with a long-range $-1/r^6$ type of interaction,
Eq.~(\ref{eq:gere}) defines the generalized scattering length and effective range
for an arbitrary $l$.
It coincides with standard definitions of scattering lengths and effective ranges
whenever they are well defined in the standard theory \cite{sch47,bla49,bet49,lev63}. 
Namely, $\widetilde{a}_{l} = a_l$ for $l=0$ and 1, and $\widetilde{r}_{el}=r_{el}$ for
$l=0$. Similar to the standard theory, $\widetilde{a}_{l}$
has a dimension of $\beta_6^{2l+1}$, and $\widetilde{r}_{el}$ 
has a dimension of $\beta_6^{-2l+1}$. The generalized scattering length
is also related to $\widetilde{A}_{sl}(\epsilon_s)$
by $\widetilde{a}_{l} = \widetilde{A}_{sl}(\epsilon_s=0)\beta_6^{2l+1}$.

With this definition of generalized scattering length, having a bound 
or quasibound state right at the threshold, 
characterized in terms of the $K^{c0}_l$ parameter by $K^{c0}_l(\epsilon=0)=0$,
always corresponds to $\widetilde{a}_l=\infty$, for any $l$.
Similarly, having a shape resonance close to the threshold,
characterized by a small and negative $K^{c0}_l$, corresponds to
having a large and negative generalized scattering length;
having a bound state close to the threshold,
characterized by a small and positive $K^{c0}_l$, corresponds to
having a large and positive generalized scattering length.
Such similarities to the $s$-wave interaction
make the generalized scattering length an easy parameter
to understand, without having to know the QDT behind it.
As examples of the generalized scattering lengths for arbitrary $l$, 
we present, in the Appendix~\ref{sec:model}, 
their analytic results for two classes of model potentials 
with $-1/r^6$ type of asymptotic behaviors.

We point out that the main utilities 
of the generalized effective range expansion,
Eq.~(\ref{eq:gere}), are (a) to define the generalized scattering length
as an alternative parameter for describing low-energy atomic interactions,
(b) to make a connection between the QDT expansion and the ERT to the degree
possible, and (c) to simplify the understanding of the QDT expansion for
peoples who are not completely comfortable with QDT formulations.
It is not meant to be a replacement for the QDT expansion.
As far as accuracy is concerned, the QDT expansion
is always more accurate. This loss of accuracy in the generalized ERT occurred 
in expanding the denominator of Eq.~(\ref{eq:cotexp}), whose full representation
would have required an infinite number of terms in the standard
ERT type of expansions. 

The procedure of expanding the denominator 
has more severe consequences in the special case of 
$\widetilde{a}_l=0$, for which 
$\widetilde{r}_{el}=\infty$ from Eq.~(\ref{eq:gr}), and
the effective range expansion, even in its generalized form here,
becomes meaningless. In comparison, the QDT expansion
remains applicable, and gives, for $\widetilde{a}_l=0$ 
[corresponding to $K^{c0}_l=-(-1)^l$], 
\begin{equation}
K^{(D)}_l \approx \frac{2\bar{a}_{sl}k_s^{2l+3}}
	{(2l+3)(2l-1)-(-1)^lk_s^2-(-1)^lw_l k_s^4} \;.
\label{eq:KDl0}	
\end{equation}	
This result also implies that $\widetilde{a}_l=0$ changes the threshold 
behavior of $K_l=\tan\delta_l$ for $l<2$.
For $\widetilde{a}_l\neq 0$ (and $\widetilde{a}_l\neq\infty$),
it is clear from the QDT expansion that the threshold
behavior for $l<2$ is determined by the $K^{(D)}_l$
term that behaves as $K^{(D)}_l\sim -\widetilde{a}_{l}k^{2l+1}$.
The threshold behavior for $l\ge 2$ is dominated by the Born term,
which behaves as $k_s^4$.
For $\widetilde{a}_l=0$, Eq.~(\ref{eq:KDl0}) means that the
threshold behavior for the $s$ wave changes from
$\tan\delta_{l=0}\sim -\widetilde{a}_{l=0}k$ to
$\tan\delta_{l=0}\sim -2\bar{a}_{sl=0}k_s^3/3$.
The threshold behavior for the $p$ wave changes
from $\tan\delta_{l=1}\sim -\widetilde{a}_{l=1}k^3$ to
being dominated by the Born term
$\tan\delta_{l=1}\sim -\pi k_s^4/35$.

Another special case for which the threshold behavior may be modified
is the case of $\widetilde{a}_l=\infty$ ($K^{c0}_l=0$), 
corresponding to having a bound or quasibound state right at threshold.
The QDT expansion gives for this special case
\begin{equation}
K^{(D)}_l \approx -\frac{(2l+3)(2l-1)\bar{a}_{sl}k_s^{2l-1}}
	{1+w_l k_s^2}-(-1)^l\bar{a}_{sl}k_s^{2l+1} \;.
\label{eq:KDlinf}	
\end{equation}	
It means, in particular, that the threshold behavior for $\tan\delta_l$ is modified
for $l\le 2$ for having a bound or quasibound state right at the threshold. 
Specifically, it is changed from $\tan\delta_l\sim -\widetilde{a}_{l}k^{2l+1}$,
for cases of $\widetilde{a}_l \neq \infty$ (and $\widetilde{a}_l \neq 0$),
to $\tan\delta_l\sim-(2l+3)(2l-1)\bar{a}_{sl}k_s^{2l-1}$ for 
$\widetilde{a}_l=\infty$. The threshold behaviors for
$l>2$ remain dominated by the Born
term ($\sim k_s^4$) even with a bound state
right at the threshold.
We note that the generalized effective range expansion would have
given $\widetilde{r}_{el}=-2\beta_6^2/[(2l+3)(2l-1)\bar{a}_l]$ and 
$\tan\delta_l\approx\delta^{(B)}_l-(2l+3)(2l-1)\bar{a}_{sl}k_s^{2l-1}$
for $\widetilde{a}_l=\infty$,  
corresponding to the lowest order term in Eq.~(\ref{eq:KDlinf}).

As a further comment on the generalized effective-range expansion,
we note that in arriving at Eq.~(\ref{eq:gr})
for the effective range,
we have assumed that the energy dependence of $K^{c0}_l$ is negligible.
This means that Eq.~(\ref{eq:gr}) is, strictly speaking, a single channel
result that will need to be modified for effective single channel problems
where the energy dependence of $K^{c0}_l$ is generally important 
(to be discussed in detail in the companion paper).
 With this limitation in mind, Eq.~(\ref{eq:gr}) does imply that 
in the case of a single channel,
for which the energy dependence of $K^{c0}_l$ is negligible, 
the generalized effective range is not an independent parameter, but can be
determined from the generalized scattering length. 
It further implies that the generalized effective range has the property of 
$\widetilde{r}_{el}>0$ for $l=0$ and $\widetilde{r}_{el}<0$ for $l\ge 1$,
again rigorous only for true single channel problems.

All QDT expansion results of previous two subsections, which were
parametrized using $K^{c0}_l$, can be written in terms of
$\widetilde{a}_{l}$ using Eq.~(\ref{eq:ga}), or equivalently,
\begin{equation}
K^{c0}_l(\epsilon=0) = \frac{1}{\widetilde{a}_l/\bar{a}_{l} -(-1)^l} \;.
\label{eq:Kc0ga}
\end{equation}
We do not give these expressions explicitly, in part to again
emphasize the following subtle, but important point. The expressions
in terms of $K^{c0}_l$ are more generally applicable because they make
no assumption about the energy dependence of $K^{c0}_l$. The
corresponding expressions in terms of $\widetilde{a}_{l}$ automatically
assumes the weak energy dependence of $K^{c0}_l$, since in effect we
are using $K^{c0}_l(\epsilon=0)$ at other energies. 
This subtlety is not an issue for true single-channel problems,
but becomes one for effective single channel problems derived from
multichannel cases (see the companion paper), for which the equations 
in terms of $K^{c0}_l$ remain applicable, but generally not those in terms of 
$\widetilde{a}_{l}$.

\section{QDT expansion for bound state energy}
\label{sec:schbsp}

The bound spectrum of a two-body, single channel system with $-1/r^6$ 
type of long-range interaction is given rigorously
by the solutions of \cite{gao98b,gao01,gao08a}
\begin{equation}
\chi^{c}_l(\epsilon_s) = K^c(\epsilon,l) \;.
\label{eq:bsp}
\end{equation}
where
\begin{equation}
\chi^{c}_l(\epsilon_s) = \frac{\tan\theta_l+
	\tan(\pi\nu/2)(1+M_{\epsilon_s l})/(1-M_{\epsilon_s l})}
	{1-\tan\theta_l
	\tan(\pi\nu/2)(1+M_{\epsilon_s l})/(1-M_{\epsilon_s l})} \;.
\label{eq:chic6}
\end{equation}
is a universal function of $\epsilon_s$ that depends only on the exponent 
of the long-range interaction and on the angular momentum $l$. 

For deriving the QDT expansion for the energy of the least-bound state 
that is close to the threshold, it is again more convenient to
rewrite Eq.~(\ref{eq:bsp}) in terms of the $K^{c0}_l$ parameter,
as
\begin{equation}
\chi^{c0}_l(\epsilon_s) = K^{c0}_l(\epsilon) \;,
\label{eq:bspc0}
\end{equation}
where
\begin{equation}
\chi^{c0}_l(\epsilon_s) = \frac{\chi^{c}_l(\epsilon_s)-\tan(\pi\nu_0/2)}
	{1+\tan(\pi\nu_0/2) \chi^{c}_l(\epsilon_s)} \;.
\label{eq:chic06}
\end{equation}
Using expansions of $\nu$, $\theta_l$, and $M_{\epsilon_s l}$,
as given by Eqs.~(\ref{eq:nuexp})-(\ref{eq:Mexp}),
we have, for sufficiently small energies or sufficiently large $l$,
\begin{equation}
\chi^{c0}_{l=0}(\epsilon_s)\approx 
	\frac{\theta_l+\pi(\nu-\nu_0)/2+\bar{a}_{sl=0}\kappa_s-\bar{a}_{sl=0}\kappa_s\theta_l} 
	{1-\bar{a}_{sl=0}\kappa_s-\bar{a}_{sl=0}\kappa_s\theta_l} \;,
\label{eq:chic06sexp}
\end{equation}
\begin{equation}
\chi^{c0}_{l=1}(\epsilon_s)\approx 
	\frac{\theta_l+\pi(\nu-\nu_0)/2-\bar{a}_{sl=1}\kappa_s^3} 
	{1-\bar{a}_{sl=1}\kappa_s^3} \;,
\label{eq:chic06pexp}
\end{equation}
\begin{equation}
\chi^{c0}_{l\ge 2}(\epsilon_s)\approx \theta_l+\pi(\nu-\nu_0)/2 \;.
\label{eq:chic06dgexp}
\end{equation}
Here $\kappa_s=(-\epsilon_s)^{1/2}$, and $\theta_l$ and $\pi(\nu-\nu_0)$
are given by Eqs.~(\ref{eq:thetal}) and (\ref{eq:qdtexp2}), respectively.

Having a bound state of angular momentum
$l$ close to the threshold corresponds to having a small and positive $K^{c0}_l$. 
The energy of such a state can be obtained by solving 
Eq.~(\ref{eq:bspc0}) perturbatively using expansions given by
Eqs.~(\ref{eq:chic06sexp})-(\ref{eq:chic06dgexp}).
This has been done in Ref.~\cite{gao04c}. We summarize the results here
using the standardized notation ($x_l$ of Ref.~\cite{gao04c} is renamed $K^{c0}_l$),
for the sake of completeness and easy reference.

For the $s$ wave, we obtained
\begin{equation}
\epsilon_{sl=0} = -\frac{1}{\bar{a}_{sl=0}^2} (K^{c0}_{l=0})^2
	\left[1+g_1 K^{c0}_{l=0}+g_2 (K^{c0}_{l=0})^2\right]+O((K^{c0}_{l=0})^5) \;,
\label{eq:es0}	
\end{equation}
where $\epsilon_{sl}$ is the $l$ wave bound state energy scaled according to
Eq.~(\ref{eq:es}), and $g_1 = 2[1/(3\bar{a}_{sl=0}^2)-1] \approx 0.9179195$,
and $g_2 = (5/4)g_1^2-2 \approx -0.9467798$.

For the $p$ wave, we obtained 
\begin{equation}
\epsilon_{sl=1} = -5 K^{c0}_{l=1}
	\left[1+h_1 (K^{c0}_{l=1})^{1/2}+h_2 K^{c0}_{l=1}\right]+O((K^{c0}_{l=1})^{5/2}) \;,
\label{eq:es1}	
\end{equation}
where $h_1 = 5^{3/2}\bar{a}_{sl=1} \approx 1.299466$, and
$h_2 = 3h_1^2/2-5\pi/14 \approx 1.410919$.

For $l\ge 2$, the energy of the least-bound molecular state is given by 
Eq.~(\ref{eq:esl}), namely the same equation that gives the position
of the shape resonance above the threshold. This comes from the fact
that Eq.~(\ref{eq:bspc0}), with $\chi^{c0}_l$ given by 
Eq.~(\ref{eq:chic06dgexp}), is the same as Eq.~(\ref{eq:shapepos}),
which determines the shape resonance positions.
The only difference is that 
while $K^{c0}_l$ is small and negative for a shape resonance close 
to the threshold, it is small and positive for a bound state close 
to the threshold.

In cases of true single channel problems, 
these equations can again be written in terms
of the generalized scattering length using Eq.~(\ref{eq:Kc0ga}).
The corresponding equations for the $s$ and $p$ waves can be found
in Ref.~\cite{gao04c}. We skip writing down these equations
explicitly to again emphasize the greater applicability of 
the equations in terms of $K^{c0}_l$. Namely, unlike the equations
in terms of the scattering length or generalized scattering length,
which assume the energy-independence of $K^{c0}_l$, the equations
in terms of $K^{c0}_l$ are, strictly speaking, applicable
even when $K^{c0}_l$ itself depends on energy. The only subtlety
here is that when the right-hand-sides of equations such as
Eqs.~(\ref{eq:es0}) and (\ref{eq:es1})
become energy-dependent, they need to be solved again to
obtain the bound state energies.

The results presented in this section have been verified in Ref.~\cite{gao04c}
through comparisons with the exact QDT results, obtained by
solving Eq.~(\ref{eq:bsp}) or (\ref{eq:bspc0}) numerically.
The result for the $s$ wave, Eq.~(\ref{eq:es0}), has also been verified by 
Derevianko \textit{et al.} \cite{der09}
and by Julienne and Chin \cite{jc08}
through independent numerical calculations.
In summary, the formula for the $s$ wave bound state energy
is applicable for $a_{l=0}>2\bar{a}_{l=0}$, corresponding, 
approximately, to $a_{l=0}>\beta_6$ or $|\epsilon_{sl=0}|<4$.
The formulas for other partial waves are applicable over
a greater range of bound state energies \cite{gao04c}.
The intermediate equations, including the expansions of the $\chi^{c0}_l$ function
as given by Eqs.~(\ref{eq:chic06sexp})-(\ref{eq:chic06dgexp}), 
and Eq.~(\ref{eq:bspc0}), which is applicable
regardless of the energy variations of $K^{c0}_l$,
will be useful in developing the analytic description
of magnetic Feshbach resonance of arbitrary $l$, 
to be presented in a companion paper.

\section{Discussions}
\label{sec:discuss}

We make here some miscellaneous comments on various aspects of the theory,
which, while not essential, may be helpful in the understanding of this
and related theories.

(a) The QDT expansion breaks naturally into two terms,
a Born term $K^{(B)}_l$, and a deviation from the Born term $K^{(D)}_l$.
This separation of $K_l=\tan\delta_l$ is also convenient for a number
of other applications including, e.g, the understanding of angular distribution
(see, e.g., Refs.~\cite{tho04,bug04}).
There are other separations of $K_l=\tan\delta_l$ possible.
For example, it can be separated into a background term and an
interference term, as suggested in Ref.~\cite{gao08a}.
One can show, however, that these two types of separations are the same
for $l>2$, as is reflected in the fact that the $K^{(D)}_l$ term
has negligible contribution to the background for $l>2$ 
[see Eqs.~(\ref{eq:KDshape}) and (\ref{eq:KDbgdg})]. 

(b) Unlike the cases of $s$ and $p$ partial waves, the generalized scattering length
for $l\ge 2$ has little effect on scattering cross sections around the threshold,
except in determining the location of the shape resonance, 
and if there is one, the cross sections around it.

(c) For true single channel problems, having an ultracold shape resonance
can only happen by accident. In this sense, the most important applications 
of a theory for ultracold shape resonance are not to the true single channel 
problems, but to the effective single
channel problems associated with Feshbach resonances, for which there will be
at least one state in the threshold region as the Feshbach resonance is tuned
around it. It is for this reason that we kept stressing the formulas that
remain applicable when $K^{c0}_l$ becomes energy-dependent, which is the
single most important difference between an effective single channel 
problem for a Feshbach resonance, and a true single channel problem. 

It is for the same reason that we did not emphasize the relationship 
of $K^{c0}_l$ for different $l$.
For a true single channel problem, the $K^{c0}_l$ parameter for different $l$
are not independent, but are related to each other through Eq.~(\ref{eq:Kc0l}),
in which $K^c$ is approximately independent of $l$ \cite{gao01}.
As a result, the $K^{c0}_l$ for the first few partial wave can all
be determined from, e.g., the $s$ wave scattering length.
This simple relationship between different $l$ breaks down for
the effective single channel problems for Feshbach resonances. 
For such intrinsically multichannel problems, the $K^{c0}_l$ parameters for 
different $l$ are still related, as the underlying short-range $K^c$ matrix
is still approximately $l$-independent. 
Their values for different $l$ can still be computed from, e.g., the singlet
and the triplet $s$ wave scattering length for alkali-metal atoms \cite{gao05a}.
Their relationship, however, becomes more complicated than Eq.~(\ref{eq:Kc0l})
and less transparent.

(d) While we have no intention of promoting one parameter over the other,
as they all have different utilities, we hope this work again illustrates 
that the scattering length, or the generalized scattering length, is
not the only parameter, or necessarily the best parameter for
characterizing atomic interaction at low temperatures.
While many of our results, such as those for resonance positions 
and binding energies, can be written in terms of the scattering length
or the generalized scattering length, such equations are more complex
than those in terms of $K^{c0}_l$, and have more restricted applicability
to the case of single channel. Furthermore, it should be clear that in 
the scattering length representation, what is important is not the
scattering length itself, but the dimensionless ratio 
$\widetilde{a}_{l}/\bar{a}_{l}$, as is also recognized by
Chin \textit{et. al} \cite{chi08} for the $s$ wave.

(e) We point out that there are a number of other successful QDT 
formulations of atomic interaction that are conceptually similar to
ours in many aspects, but based on numerical reference functions 
\cite{mie84a,mie84b,jul89,bur98,mie00,rao04}.
Without incorporating at least some ingredients of an analytic solution,
numerical solutions generally run into difficulty for very large scattering
lengths, namely when there is a state very close to the threshold.
This is true for the $s$ wave \cite{jul08}, 
and more so for higher partial waves.

\section{Conclusions}
\label{sec:conclude}

In conclusion, we have presented an analytic description of atomic interactions
at ultracold temperatures for the case of single channel.
In particular, a QDT expansion for scattering has been developed that
is applicable to all angular momentum $l$, with or without the presence of
ultracold shape resonances. Using the QDT expansion, we have developed 
a fully analytic characterization of ultracold shape resonances in 
terms of its position, width, and background.
We have also introduced a generalized scattering length that can be used 
as an alternative parameter to
characterize atomic interaction at low temperatures, and discussed the 
changes of threshold behaviors in
the special cases of zero and infinite generalized scattering lengths.
The analytic formulas derived make it possible for
an accurate description of atomic interaction in the ultracold
regime without having to know the details of the QDT or resort to
numerical calculations.

The generalities of the results of this work will further manifest
themselves in a companion work on atomic interaction around a
magnetic Feshbach resonance, which is necessarily multichannel in
nature \cite{tie93,koh06,chi08}. 
We will show that such a multichannel problem can be rigorously reduced to
an effective single channel problem, to which most of our results
here remain applicable. The key
difference will be that the effective $K^{c0}_l$ parameter
becomes energy dependent. It is this energy dependence that leads
to deviations from the single channel universal behaviors 
that correspond to the results here specialized to an energy-independent
$K^{c0}_l$. 
 
\begin{acknowledgments}
I thank Paul Julienne for helpful discussions and communications.
This work is supported by the National Science Foundation under 
the Grant No. PHY-0758042.
\end{acknowledgments}

\appendix*
\section{Analytic results of $K^{c0}_l$ and generalized scattering 
	lengths for two type of model potentials}
\label{sec:model}

In a previous work \cite{gao03}, we have derived, for two classes 
of model potentials, the analytic results of $K^c(\epsilon=0,l)$ and 
the number of bound states $N_l$ for an arbitrary $l$.
One class, denoted by HST$n$, is of the type of a hard-sphere with
an attractive tail:
\begin{equation}
V_{\text{HST}n}(r) = \left\{ \begin{array}{lcl}
	\infty & , & r \le r_0 \\
	-C_n/r^n & , & r>r_0 \end{array} \right. \;.
\label{eq:HST}
\end{equation}
The other, denoted by LJ$n$, is of the type of Lennard-Jones LJ$(n,2n-2)$: 
\begin{equation}
V_{\text{LJ}n}(r) = -C_n/r^n+C_{2n-2}/r^{2n-2} \;,
\label{eq:LJn}
\end{equation}
which corresponds, in particular, to a LJ(6,10) potential for $n=6$.

For HST$n$ potentials, we have shown that the $K^c$ parameter at
zero energy is given by \cite{gao03}
\begin{equation}
K_{\text{HST}n}^c(0,l)
	= -\frac{J_{\nu_0}(y_0)\cos(\pi\nu_0/2)-Y_{\nu_0}(y_0)\sin(\pi\nu_0/2)}
		{J_{\nu_0}(y_0)\sin(\pi\nu_0/2)+Y_{\nu_0}(y_0)\cos(\pi\nu_0/2)} ,
\end{equation}
where $\nu_0=(2l+1)/(n-2)$, $J$ and
$Y$ are the Bessel functions \cite{abr64}, and $y_0=[2/(n-2)](\beta_n/r_0)^{(n-2)/2}$,
in which $\beta_n = (2\mu C_n/\hbar^2)^{1/(n-2)}$ is the length
scale associated with the $C_n/r^n$ type of potentials.

For LJ$n$ potentials, we have derived \cite{gao03}
\begin{equation}
K_{\text{LJ}n}^c(0,l) = \tan(\pi\nu_0/2)
	[1+h_l(z_0)][1-h_l(z_0)]^{-1} ,
\end{equation}
where $z_0=(\beta_n/\beta_{2n-2})^{n-2}/[2(n-2)]$ and
\begin{equation}
h_l(z_0) = z_0^{\nu_0}\frac{\sin\pi(z_0+1/2-\nu_0/2)\Gamma(z_0+1/2-\nu_0/2)}
	{\sin\pi(z_0+1/2+\nu_0/2)\Gamma(z_0+1/2+\nu_0/2)} \;.
\end{equation}

Substituting these results into Eq.~(\ref{eq:Kc0l}),
We obtain
\begin{equation}
K^{c0}_l(\epsilon=0) = -J_{\nu_0}(y_0)/Y_{\nu_0}(y_0) \;,
\end{equation}
for HST$n$ potentials, and 
\begin{equation}
K^{c0}_l(\epsilon=0) = \frac{\sin(\pi\nu_0)h_l(z_0)}{1-\cos(\pi\nu_0)h_l(z_0)} \;,
\end{equation}
for LJ$n$ potentials.
Both results are applicable for arbitrary $n$ and $l$.

Specializing to the case of $n=6$, these results, 
combined with QDT for $n=6$ \cite{gao98b,gao01,gao08a},
provide an accurate description of the scattering and bound state properties
for these model potentials over a wide range of energies around the
threshold. 
They also give us the analytic results of the generalized scattering
length for an arbitrary $l$.
From Eq.~(\ref{eq:ga}), we have, for $n=6$,
\begin{equation}
\widetilde{a}_{l} = \bar{a}_{l}\left[(-1)^l-Y_{\nu_0}(y_0)/J_{\nu_0}(y_0)\right] \;,
\label{eq:gaHST6}
\end{equation}
for HST6 potential, where $\nu_0=(2l+1)/4$ and $y_0=(\beta_6/r_0)^2/2$,
and 
\begin{equation}
\widetilde{a}_{l} = \frac{\bar{a}_{l}}{\sin(\pi\nu_0)h_l(z_0)} \;,
\end{equation}
for LJ6 potential, where $z_0=(\beta_6/\beta_{10})^4/8$.
The generalized effective range can be derived from these results
using Eq.~(\ref{eq:gr}).
If one specializes Eq.~(\ref{eq:gaHST6}) to $l=0$, one recovers,
for the HST6 potential, the result of Gribakin and Flambaum \cite{gri93} 
for the $s$ wave.

The analytic results of this Appendix are useful for designing model 
potentials to investigate universal properties not only in two-body \cite{gao03,gao04b},
but also in few-body and many-body quantum systems \cite{gao03,gao04a,kha06}.
They can also be used to check the accuracies of
various numerical techniques and methods.
Last but not the least, they give an explicit illustration of 
one of the important properties of $K^c$ and related parameters, i.e.,
$K^c(\epsilon, l)$, and therefore $K^{c0}_l(\epsilon=0)$ and $\widetilde{a}_l$,
are meromorphic functions of $l$, namely, they are analytic functions of
$l$ with only simple poles \cite{gao08a}. 

\bibliography{bgao,twobody,manybody}

\end{document}